# Development of an IoT Based Sleep Apnea Monitoring System for Healthcare Applications

**Abdur Rab Dhruba[1], Kazi Nabiul Alam[1], Md Shakib Khan[1], Sami Bourouis[2] and Mohammad Monirujjaman Khan[1*]**

[1]Department of Electrical and Computer Engineering, North South University, Bashundhara, Dhaka-1229, Bangladesh

[2]Department of Information Technology, College of Computers and Information Technology, Taif University, P.O. Box 11099, Taif, 21944, Saudi Arabia

*Corresponding Author: Mohammad Monirujjaman Khan. Email: monirujjaman.khan@nortsouth.edu

**Abstract:** Sleep is an essential and vital element of a person's life and health that helps to refresh and recharge the mind and body of a person. The quality of sleep is very important in every person's lifestyle, removing various diseases. Bad sleep is a big problem for a lot of people for a very long time. People suffering from various diseases are dealing with various sleeping disorders, commonly known as sleep apnea. A lot of people die during sleep because of uneven body changes in the body during sleep. On that note, a system to monitor sleep is very important. Most of the previous systems to monitor sleeping problems can't deal with the real time sleeping problem, generating data after a certain period of sleep. Real-time monitoring of sleep is the key to detecting sleep apnea. To solve this problem, an IoT based real-time sleep apnea monitoring system has been developed. It will allow the user to measure different indexes of sleep and will notify them through a mobile application when anything odd occurs. The system contains various sensors to measure the ECG, Heart Rate, Pulse rate, Skin response and SpO2 of any person during the entire sleeping period. This research is very useful as it can measure the indexes of sleep without disturbing the person and can also show it in the mobile application simultaneously with the help of a Bluetooth module. The system has been developed in such a way that it can be used by every kind of person. Multiple analog sensors are used with the Arduino UNO to measure different parameters of the sleep factor. The system was examined and tested on different people's bodies. To analyze and detect sleep apnea in real-time, the system monitors several people during the sleeping period. The results are displayed on the monitor of the Arduino boards and in the mobile application. The analysis of the achieved data can detect sleep apnea in some of the people that the system monitored, and it can also display the reason why sleep apnea happens. This research also analyzes the people who are not in the danger of sleeping problems by the achieved data. This paper will help everyone learn about sleep apnea and will help people detect it and take the necessary steps to prevent it.

## 1 Introduction

Sleep Apnea is a disorder which is parallelly connected to the human respiratory system and our brain. It indicates a problem with breathing and respiration obstacles during sleep. There are two kinds of sleep apnea: Obstructive Sleep Apnea (OSA) and Central Sleep Apnea (CSA). Sometimes they both take place at the same time. That can be said as Complex Sleep Apnea, but it's too rare to be called an apnea type. Obstructive Sleep Apnea (OSA) is basically upper airway congestion. It occurs due to relaxation of our throat muscles and lack of oxygen passing through our nasal-throat passage that causes irritation in breathing and sometimes results in serious complications. Central Sleep Apnea is less popular but more alarming than OSA. The human brain is the key maintenance center of the whole body, as every organ and system runs on its directed signal or instruction, and its Central Sleep Apnea comes into major exposure. Central Sleep Apnea occurs when the brain fails to send the necessary instructions or signals to the system which controls our breathing and respiration. Here, neurons fail to transmit signals to the breathing muscle, which pauses the breath for a good amount of time, maybe near 10 seconds [1]. And in rare cases, they can both take place, which is more than an emergency medical treatment issue, and alarming also.

Medical and clinical research are going on regarding this issue, and some alarming scenarios are keeping us alert. Sleep apnea is responsible for a wide range of physical complications and diseases, including strokes, hypertension, cardiac abnormalities, depression, and others. [2]. According to research, around 3 to 7% of men and 2% to 5% of women are suffering from sleep apnea. That sums up approximately more than 100 million people in the world, including adolescents. Interestingly, 80% of apnea cases remain undiagnosed. Sleep apnea (OSA) affects approximately 1 to 4% of children aged 2 to 8, with 20% of them snoring. It varies in different parameters, like who is affected, who is not [3], [4]. Some complications or risk factors related to this problem identifying process are: excessive weight or obesity, being men (2-3 times higher risk), alcohol consumption, smoking, family history, neck circumference, nasal congestion, medical conditions like high blood pressure, type-2 diabetes, lung or other respiratory diseases etc. Patients with diseases like Parkinson's and after strokes are at risk of suffering from central sleep apnea. Sleep Apnea can cause a good number of health complications, and death in the very worst case. Loud snoring, fatigue, drowsiness, weakness in the body, lack of concentration are the common symptoms of sleep apnea. High blood pressure caused by a lack of sleep and a low oxygen level may also increase the risk of a heart attack. Apnea patients are three times more likely to have a stroke. Lungs can be affected even more dangerously when the SpO2 level is reduced. Lung disability caused by a lack of oxygen is a very common but very concerning complication of sleep apnea. 43% of people suffering from mild sleep apnea had 'hypertension'. According to research, this apnea problem causes 15% of traffic accidents and costs approximately 1000 lives in the United States. Over 38, 000 people in the USA die every year due to direct and indirect effects of sleep apnea, mostly affected by cardiovascular complexities [5]. So, this is a matter of awareness as it's not being focused on yet or getting that much exposure, but we'll try to identify it with ease with our research so that people don't remain undiagnosed and can have a comfortable life with no life risk.

The very basic parameters of health which are related to sleep apnea include: AHI (Apnea-Hypopnea Index), Heart Rate (BPM), Blood Oxygen Saturation (SpO2), Body Mass Index (BMI), Sleeping time, REM, Blood Glucose, Age, Blood Cholesterol, and many more [6]. In different research, it has been seen that such parameters are covered on the basis of medical research. In most cases, they come up with a study of medical data of a particular or group of patients and they analyze the results. Devices are mostly used: Polysomnography (PSG), Electrocardiogram (ECG) and some clinical devices like this. These devices are too costly to use for personal use. Some advanced and fruitful research happened earlier based on sleep apnea. In [7], wearable e-textile sensors used an IoT approach to collect real-time data on sleeping habits and respiratory rate. Same concept is used in [8] as they use signals of breathing collected from the sensor-based mattress, which are generally used in operation theatres. A Polyvinylidene Fluoride (PVDF) sensor is used with a polysomnography (PSG) device to measure time-to-time AHI data [9]. Researchers used the 'Peripheral Arterial Tonometry' (PAT) method to detect sleep apnea perfectly. They developed the WatchPAT device that records the pulse wave on a finger and derives sleep and sleep apnea features [10]. In [11], a 3D camera was being used to check the movement of abdominal muscles which they compared with the OSA detection process. The sound detection method used in [12] is one that allows for precise measurement of sounds with very high frequencies. Besides, so much research has taken place with the analysis of real-time ECG monitoring and pre-reserved ECG data. Machine Learning techniques were implemented. In [13], an IoT device was built that extracts ECG data from the user and evaluates the result from the classification model (SVM) based system, which they named the "Apnea MedAssist Service". In [14], [15], [16], researchers implemented Machine Learning and Deep Learning techniques to detect apnea from medical data.

From earlier research, it is observable that researchers measured real-time parameters of sleep apnea, which they analyzed afterwards. But there is less use of parameters of sleep apnea like: SpO2, Heart Rate, AHI, Sleep Time or such important parameters being measured at a time. In particular, they measured data and analyzed it with advanced tools. Besides, in earlier research, these values were measured overnight and then they got the data, but in this system the value will be measured in real time. If any complications arise, someone concerned may be notified via IoT device. For example, if "Oxygen Saturation" falls, this is a severe medical emergency, and with a possible device, the attendant, doctor, or concerned personnel can get instant signals from this individual patient. Due to sleep apnea, people may have a stroke, may have cardiac arrest, and may have heart failure or such complications. Real time measurement of these apnea parameters is a must to reduce such risks, as it may cause death. Another motivation is that regular apnea screening devices are too costly to afford. They are mostly used in medical institutes for medical conditions or research purposes, but a cost-effective device is needed so that the general public can buy this and screen their apnea issues while staying at home, with no need to rely on costly clinical monitoring. Our research goal is to solve these issues and saving human lives is the ultimate goal of this study.

An IoT based sleep apnea monitoring system for healthcare applications is developed in this study. In this research, the system will monitor the parameters of Sleep Apnea on a real time basis and will try to monitor the values of these as efficiently as possible with diverse parameters. The main parameters covered in this research are: SpO2, Heart Rate, ECG, Skin response and Sound intensity to be measured at a time. An application is being made that will show the real time data of these parameters so that the patient's attendant can get an instant and continuous update of the patient's health.

The rest of this paper is organized as this pattern: Section 2 describes the methods and tools implemented in our research. All the modules (hardware and software) and their workflows are being discussed here. Section 3 demonstrates the results and outcomes we got from our system and their broad qualitative analysis as well. In section 4, the whole concept of this research is summarized, and we also discussed how important and effective it is to monitor sleep apnea on a real-time basis, and we concluded the section by discussing our ambitions and motivation for this work for the sake of lives.

## 2 Methods and Materials

### 2.1 Methodology Statements

This section discusses the methods, components and paths that are used to fulfill the goal. The aim of the system is to monitor the entire sleeping period of a patient with sleeping disorders. This microcontroller based sleep apnea monitoring system for sleeping disorder patients is combined with three different layers. The main layer is a microcontroller unit which connects the input layer and the output layer. In the input layer, it is combined with four different sensors which will provide the analog signal to the Arduino UNO to measure the different indexes of sleep condition. The output layer is combined with two parts, including the serial monitor of the Arduino UNO and a mobile application to display the digital data converted by the microcontroller.

### 2.2 Outline of System

A block diagram showing the full system in Fig.1. The system consists of input, output and a microcontroller board Arduino UNO shown in Fig. 2. The Arduino board, which is also connected to the output layer, combined with the serial monitor of the Arduino board, and a MIT App Inventor based mobile application connected to a Bluetooth module, show the converted digital data to the viewer.

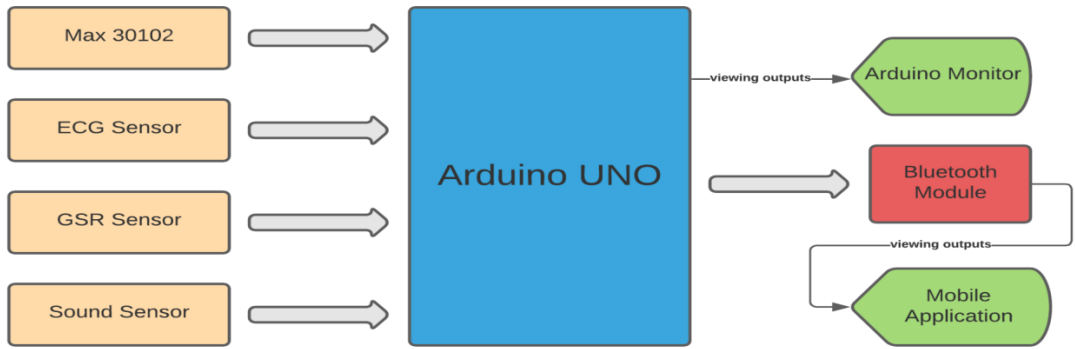

**Figure 1:** Full system block diagram by which the system works successfully.

Fig.1 shows the basic workflow of the system. The sensors provide the data to the Arduino UNO simultaneously, and the Arduino UNO passes the converted digital data to the Arduino IDE (Integrated Development Environment)'s serial monitor and also to the Mobile Application through the Bluetooth module at the same time.

## 2.3 Modules and Materials

The system is integrated with different kinds of components that are doing different tasks in the system. Some are for the input, some are for the output, and some are used in the system to create a bridge between the inputs and outputs.

### 2.3.1 Arduino UNO

The Arduino UNO, shown in Fig. 2, is the system's major component, and it's based on the VR Microcontroller Atmega328. This programmable microcontroller has the capacity to connect to other sensors or computers, allowing it to be used in many projects. It has 2KB of SRAM (Static Random Access Memory) and 32KB of flash memory, 13KB of which is utilized to store the set of instructions in the form of code. It also includes a 1KB EEPROM (Electrically Erasable Programmable Read-only Memory).

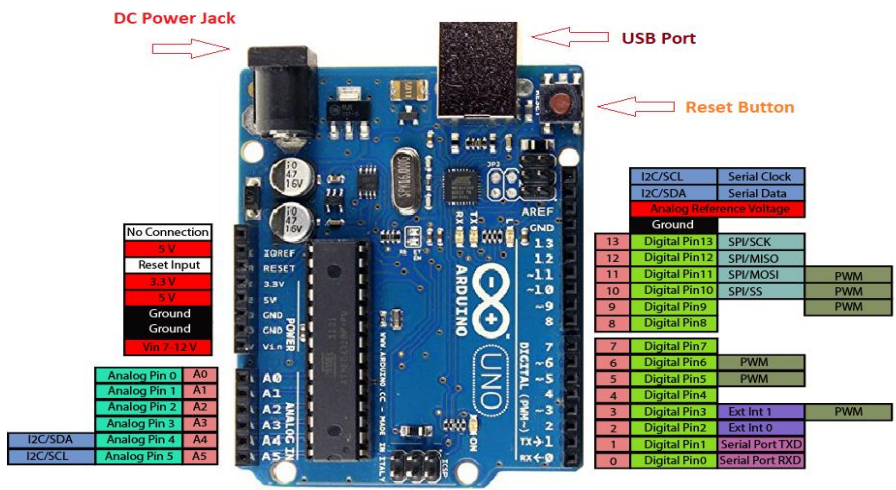

**Figure 2**: Arduino UNO Pinout [17]

This Arduino board has a total of 30 connections, with 14 digital pins and 6 analog pins for external connection. The A0 to A5 analog pins are used to receive analog data from external devices such as analog sensors. On the board, there are various digital and analog input and output pins that operate at 5V. These pins have conventional operational current ratings of 20 to 40 milliamps. The DC power jack which may provide a voltage ranging from 7V to 20V, or the USB connected to an external device can provide a 5V voltage. Data transmission is the key for IoT devices. To receive, transmit data, and maintain serial communication, two pins called Pin 0 (Rx) and Pin 1 (Tx) work simultaneously. The Rx pin receives data, whereas the Tx pin transmits data. Serial communication can also be done by other Input/Output (I/O) pins of the board. The Serial monitor of the Arduino IDE software is used to send or receive text data from the board. It is used to display the output data from the Arduino board.

*2.3.2 Sensors*

There are several sensors to monitor the patient's condition during sleep. To monitor and analyze sleep apnea, the system contains a few sensors that will fulfill the major goals. The heart rate module, SpO2, Pulse sensor, ECG sensor, Galvanic Skin Response (GSR) sensor and Sound sensor to monitor the snoring sound of the patient during sleep have been used in the system. These six sensors will monitor the patient's during sleep and will provide analog data to the Arduino UNO.

*2.3.2.1 Heart Rate Pulse Sensor*

Difficulties with breathing and sleep apnea problems are very much related to the heart rate of the patient. The system contains a heart rate pulse sensor to monitor the heart rate of the patient throughout the sleeping period. A high heart rate is one of the major risk factors for sleep apnea. The normal heart rate of a healthy person is in the range of 60-100 beats per minute (bpm). The heart rate varies from person to person. More physically active people usually have a lower heart rate than less active people. A higher and unusual change in heart rate is a major sign of sleeping disorders. Fig. 3 shows the heart rate pulse sensor.

The heart rate during sleep is usually lower than at other times of the day. The average heart rate of a person during sleep is usually around 60-80 bpm, as shown in Fig. 4, but a healthy and fit person has a lower heart rate in the range of 50-60 bpm during the sleeping period.

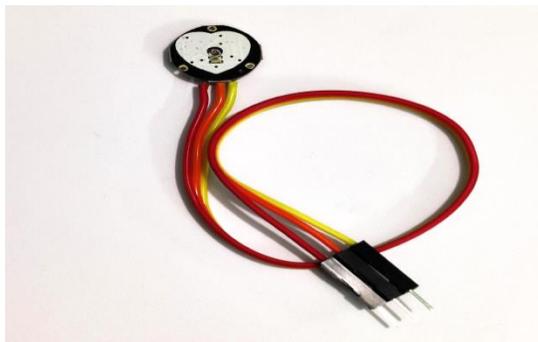
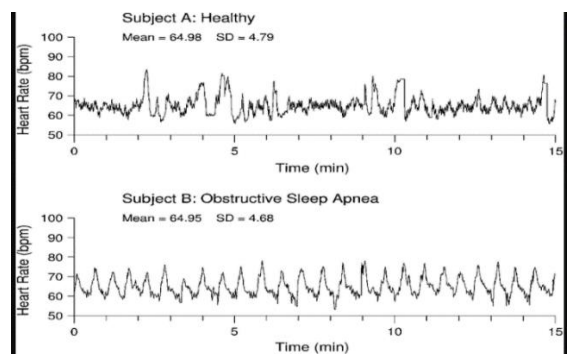

**Figure 3:** Heart Rate Pulse Sensor    **Figure 4:** Heart Rate during Sleep [18]

An unusual change in heart rate is one of the symptoms of sleep apnea. A lower heart rate can occur due to a lack of oxygen supply in the body. It could happen when the breathing of a person stopped due to any unusual reason. A higher heart rate is also a result of sleep apnea. Because of hypertension and stress, a higher heart rate can occur. And it usually hampers the usual sleep of a person. It is one of the main reasons for heart failure during sleep.

The system will include a heart rate sensor to continuously monitor the person's heart status. This is an analog sensor, and the sensing element of the sensor will be attached to the person's body. This will send the analog signal to the Arduino through the analog pin. The heart pulse sensor is depicted in Fig.3. The data pin of the sensor is connected with one of the analog pins of the Arduino for data transmission. This sensor operates on a voltage of 5V and 4mA current.

*2.3.2.2 AD8232 ECG Sensor*

An electrocardiogram, or ECG, is very important in monitoring sleep apnea. An ECG provides information about the heart rate and the rhythm of the heart. It also shows the unusual change in the heart rate. It can provide the state of the heart during the sleeping period, whether there is any kind of enlargement of the heart occurring due to hypertension, and can also detect myocardial infarction. The ECG varies person to person for several reasons. A physically active person has a more stable ECG than an inactive person's ECG. The resting ECG is different from the stress or exercise ECG. Fig. 5 shows the AD8232 ECG sensor used in this research. Fig. 6 given below will show the ECG values during sleep.

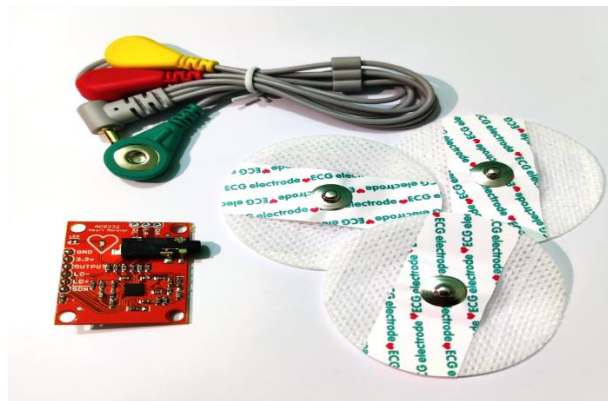
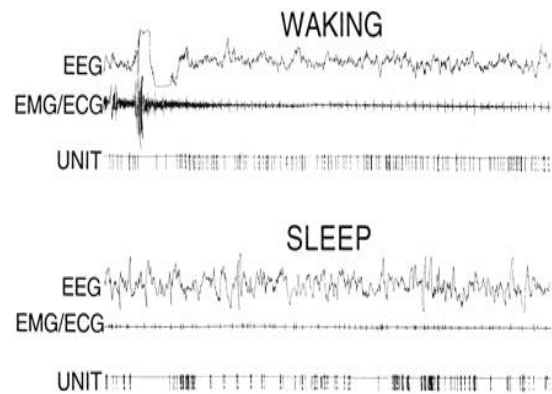

**Figure 5:** AD8232 ECG Sensor          **Figure 6:** ECG Curves

The system will contain an AD8232 ECG analog sensor to monitor the ECG of a person shown in Fig. 5. That sensor will be connected to the body and pass an analog signal to the Arduino to convert the signal into a digital signal. This sensor is connected with the Arduino via five pins. Among them, two are for powering up the sensor. The input voltage of the sensor is 3.3V. Other pins are the data pin, which is connected to the analog pin of the Arduino, and others are the lead pins LO+ and LO-. This lead pin analyses the electrode which is connected to the body and gives the data to the Arduino. These ECGs can be excessively noisy at times.

*2.3.2.3 Max 30102 Finger Oximeter Heart Rate Module SpO2*

The amount of oxygen circulating in the blood is referred to as the blood oxygen level. Oxygen distribution throughout the body is an indicator of detecting a healthy body and an unhealthy body. The blood oxygen level or oxygen saturation level indicates how equally oxygen is distributed in the body from the lungs to the cells. Problems in the lungs are one of the major reasons for low oxygen saturation. Breathing problems can cause a low supply of oxygen to the lungs and the body. During OSA, the breathing rate usually falls. This causes a lower supply of oxygen to the body. That change affects the rate of oxygen saturation of blood. A healthy person's oxygen saturation, which is also referred to as SpO2, is 95% or higher. A sleep apnea patient usually has a lower SpO2 value which is usually around 90%. Lower SpO2 can cause hypoxemia. It also indicates chronic lung disease. Patients need external oxygen supply if the oxygen saturation values fall from these threshold values. Fig. 7 shows the oxygen saturation during a sleep apnea situation.

To measure the blood oxygen level, the system contains a sensor-based pulse oximeter. The MAX30102 sensor will be used in the system to monitor the oxygen saturation in blood shown in Fig. 8. This sensor combines two Light Emitting Diodes (LED), a photo detector, optimized optics, and low-noise analog signal processing to detect Pulse Oximetry. Fig. 9 given below shows the system block diagram of MAX30102.

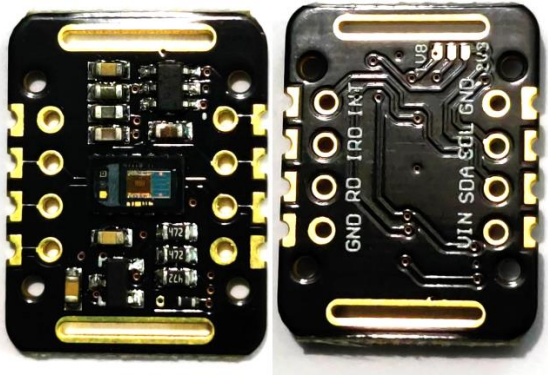
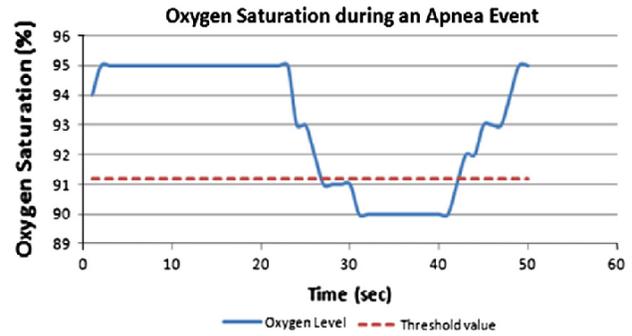

**Figure 7:** Max30102 SpO2 Sensor

**Figure 8:** SpO2 Level during Sleep Apnea [20]

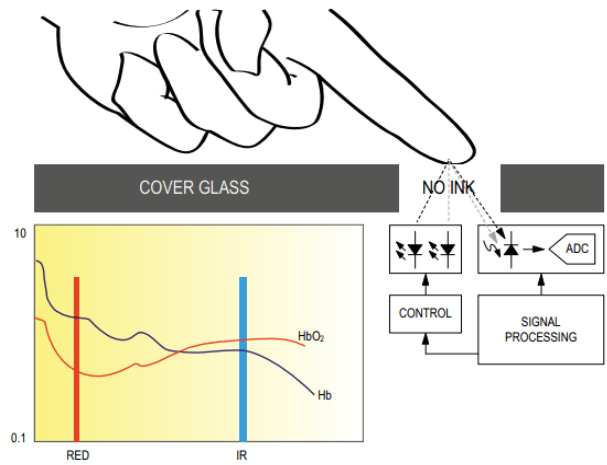

**Figure 9:** System Block Diagram of MAX30102 [21]

The main mechanism behind measuring the SpO2 value by this sensor is to detect the ratio of blood-carrying hemoglobin and blood-not-carrying hemoglobin. This sensor can measure the SpO2 value from the fingertip of a person. It operates on a voltage ranging from 3.3V to 5V and senses the data from the person's body and passes it to the Arduino to convert the data.

*2.3.2.4 Galvanic Skin Response (GSR) Sensor*

The galvanic skin response or GSR, also known as emotional arousal, refers to variations in sweat gland activity that reflect the degree of our emotional state [22]. It indicates the electro dermal activity of a person's body. It is the result of sweat glands in the skin being activated by the autonomic nervous system. The skin response is different at different times of the day. It also changes with the state of mind of a person. A stressful person tends to have a lower skin response than a joyful person. The GSR value also tends to be lower during sleep. Sweating hands is also a factor in skin response. Excessive sweating is one of the main results of hypertension. It reflects the lower value of skin response.

The value of GSR is measured by the conductance of the skin. The equation to measure skin response is:

$$\text{Conductance} = (1/\text{ Resistance}) \tag{1}$$

The value of skin response is measured in micro-Siemens. The average value of the GSR of a normal person is in the range of 250 µS to 450 µS. Sleeping GSR usually falls as the period of sleep increases. But any unusual change in skin response can cause major issues. A sudden drop in response is a symptom of sleep apnea as it can occur with sudden emotional state changes caused by stress or hypertension. Fig. 10 shows the GSR sensor module and Fig. 11 shows the changes of electro dermal activity during sleep.

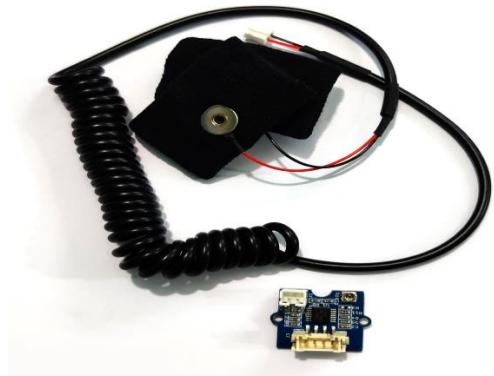
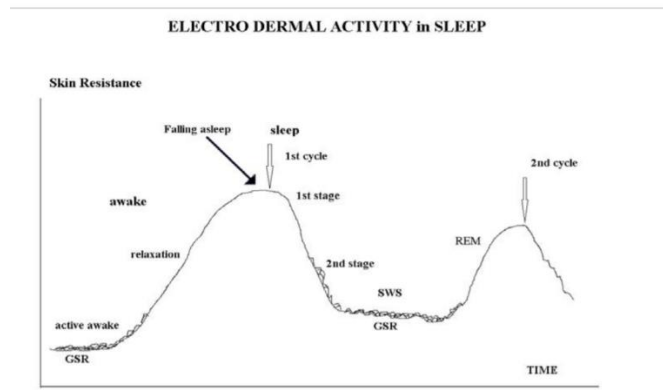

**Figure 10:** GSR Sensor Module        **Figure 11:** ED Activity during Sleep [22]

To measure the galvanic skin response during sleep and monitor sleep apnea, the system contains a wearable analog sensor. The Galvanic Skin Response sensor enables the system to monitor sweat gland activity, which is linked to emotional arousal. This sensor is easily wearable on the finger to measure the GSR value. The sensor passes the value to the Arduino to convert it into digital data, and later it passes to the mobile application.

*2.3.2.5 Sound Sensor*

Snoring is one of the major symptoms of Obstructive Sleep Apnea or OSA. It indicates the problem with breathing. Snoring is a raspy or loud sound produced by air passing through relaxed tissues in the throat, causing the tissues to vibrate while you breathe. From a health standpoint, OSA-related snoring is more concerning. To detect OSA, monitoring of snoring sound and frequency is important. An unusual change in snoring replicates the intensity of the sleep apnea problem. Fig. 12 shows the sound sensor module.

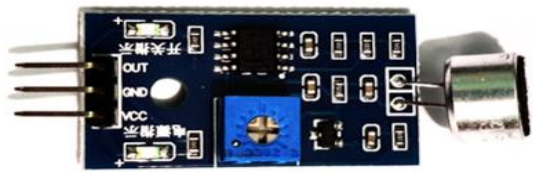

**Figure 12:** Sound Sensor

To measure the intensity of the snoring, the system will contain an analog sound sensor that will measure the snoring sound and determine the snoring during the sleeping period. The sound sensor is a compact board with a microphone and some processing circuitry for translating sound waves into electrical signals. In the system, the sensor is connected to the Arduino via the analog pin of the Arduino. Arduino gets the analog data from the sensor and translates it into digital for the mobile application.

*2.3.3 Bluetooth Module*

Serial communication is the key to this IoT-based paper. For that, the system contains a Bluetooth module HC-05. This Bluetooth module is the gateway between the Arduino Uno and android application. Operated in two moods, this Bluetooth module sends or receives data to another device in one mode, and another mode is working in AT Command mode to set the device's settings as default. After pairing with the Bluetooth device, the digital data will be visible in the mobile application for the user. Fig.13 shows the pin out of the HC-05 Bluetooth module.

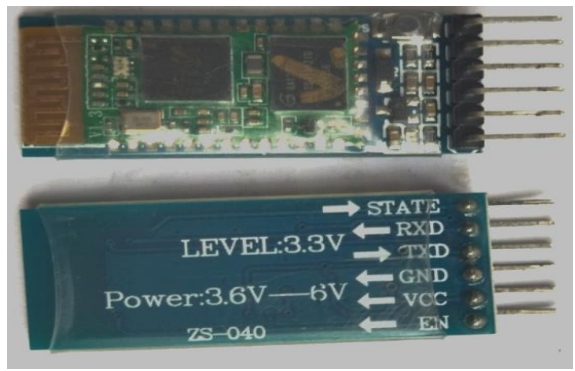

**Figure 13:** Bluetooth Module

The devices operate in a voltage range of 4V to 6V and a current of 30mA. The TX(Transmit) and RX (Receive) pins of the Bluetooth module operate serial communication. To transmit serial data, the TX pin operates, and the RX pin works to receive data from the microcontroller. The TX pin of the Arduino is connected to the RX pin of the Bluetooth module, and the RX pin of the Arduino is connected to the TX pin of the Bluetooth module.

In the system, after pairing with the Bluetooth module, the mobile application, digital data is transmitted to the app from the Arduino.

*2.3.4 Components and Cost*

The system is quite cost-effective for detecting sleep apnea. Tab. 1 shows the list of components required to build the system and the cost of it. The only cost of the system is for the hardware components, especially the sensors. Some of the sensors are priced at more than 850 Bangladesh Taka (approximately 10 US dollars). The total cost of the system is 4570 Taka ($53.76). Although the system is quite cost-effective, it is very useful for monitoring sleep apnea. One patient at a time can utilize this device, and can be used for multiple times as per the requirements.

**Table 1:** List of Components and Cost

| Name of Components | Units | Cost (BDTaka) |
|---|---|---|
| Arduino Uno | 1 | 400 Taka ($4.7) |
| Heart rate Sensor | 1 | 300 Taka ($3.53) |
| AD8232 ECG Sensor | 1 | 950 Taka ($11.18) |
| MAX30102 | 1 | 850 Taka ($10) |
| Sound Sensor | 1 | 150 Taka ($1.76) |
| GSR Sensor | 1 | 1200 Taka ($14.12) |
| HC-05 Bluetooth Module | 1 | 300 Taka ($3.53) |
| Breadboard | 1 | 90 Taka ($1.05) |
| Jumper Wires | 1 Box | 130 Taka ($1.53) |
| Others | - | 200 Taka ($2.35) |
| Total = | | 4570 Taka ($53.76) |

### 2.3.5 MIT App Inventor 2

This system contains a mobile application to show the real-time data transferred from the Arduino via Bluetooth. This application shows all of the parameters related to sleep apnea continuously. In the system, the name of the application is "Sleep Apnea Monitoring Device". This mobile application was created using MIT App Inventor 2 [23]. It allows the user to create an application and also has the functionality to create a gateway between the hardware devices. This mobile application shows the digital data converted by Arduino UNO. To do that, the Bluetooth module HC-05 is connected to the Arduino. After pairing with the mobile and connecting to the application, the values become visible in the mobile application. The layout of the application that we made in the MIT App Inventor is shown in Fig. 14.

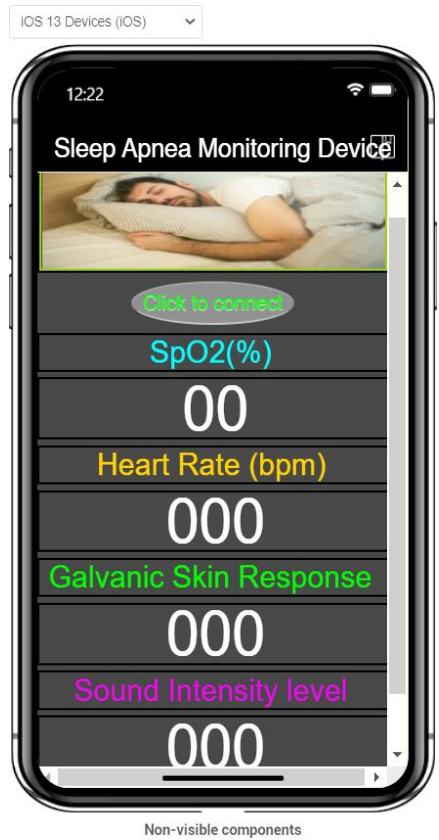

**Figure 14:** Sleep Apnea Monitoring App in the MIT App Inventor Website during Development.

*2.3.6 Full System Review and Working Process Flow-chart*

The system is combined with a few different steps. Every step has a different task to do and both make it useful for the user. Our whole hardware setup we showed in Fig.15. The programmable Arduino UNO program combines those different steps. A flow-chart of the Arduino program is shown in Fig. 16. The Arduino program is combined with different functions for the different sensors and Bluetooth modules that connect the Arduino UNO with the mobile application.

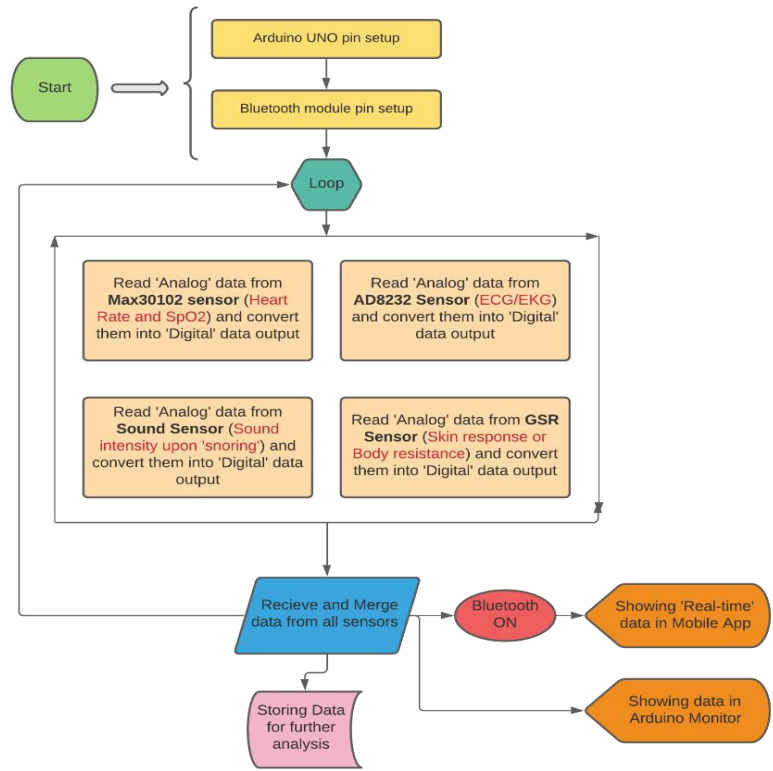

**Figure 15**: Whole System working Flow-chart

The system's work flow started from the Arduino IDE after uploading the code to the Arduino UNO. The code contains the setup of the sensors, the working process of the sensors, and the Bluetooth module. The main function of the code is controlled by a loop which allows the system to display the data every second. The loop controls the organization of the Max30102 module, ECG module, the sound sensor, and the GSR sensor. Also, the data conversion and transmission parts are performed in this section, and the data passed to the serial monitor and mobile application happens in this part. After pairing the Bluetooth module with the mobile, the data is displayed on the serial monitor and also stored for detailed analysis.

*2.3.7 Diagrams Drawn Using Software Showing the Layout of the System*

The circuit diagram with all the sensors and Arduino Uno is designed using circuito.io [24], an online IoT design platform. This website allows you to design mostly Arduino-based devices. The hardware part of the system was created on the basis of this design. Circuit diagram of the system is given in Fig. 16.

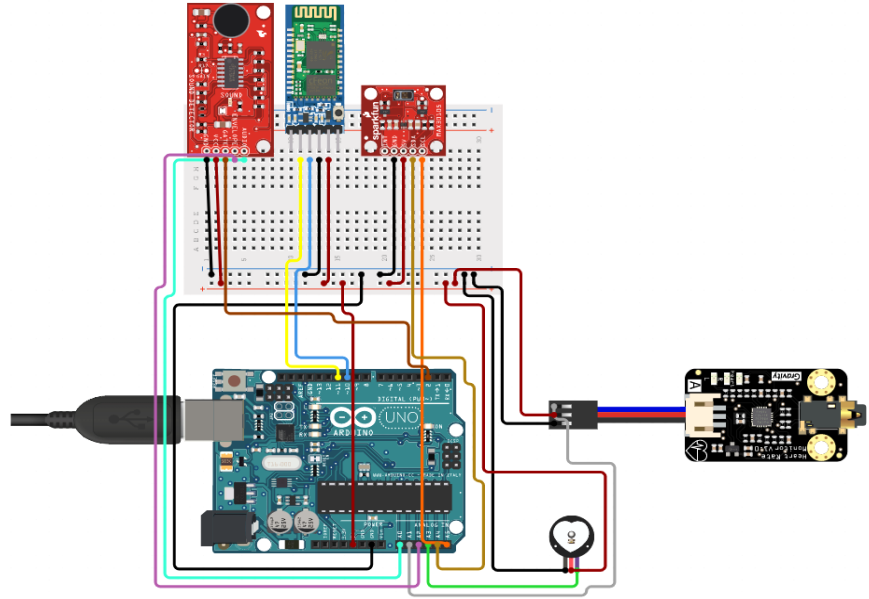

**Figure 16:** Circuit Diagram of the Full System

The hardware setup of the study was done by following the diagram shown in Fig. 16. The sensors are connected with the Arduino UNO in the way displayed in the diagram. The data pin of all of the sensors is connected to the analog pin of the Arduino individually, as shown in the figure, which is also connected to the positive and negative pins of the breadboard to get power. The Bluetooth module is connected to two of the I/O pins of the Arduino to continue serial communication.

## 3 Result and Analysis

In this investigation, it has been discovered and managed to deconstruct a number of outcomes. In the next sections, the results and observations will be discussed in more detail.

### 3.1 Patient list of Examination

To detect sleep apnea in patients, the system monitored different people from different age groups and different health conditions. The system examined both male and females and got to see quite remarkable differences. The device attempted to examine a patient with various health issues because previous health reports and disease history are one of the factors contributing to sleep apnea. The focus was on the body condition of the person before monitoring sleep apnea. The IoT device monitors the patients and provides us the results against all of the parameters. The list of persons examined by the system is given below in Tab. 2 with their age group and gender.

**Table 2:** Lists of people that has been examined.

| List of Person | Age Group | Gender | Body Status |
| --- | --- | --- | --- |
| Person-1 | 5-17 | Male | Normal |
| Person-2 | 18-35 | Male | Overweight, Excessive sweating Problem |
| Person-3 | 18-35 | Male | Physically Fit (Athlete) |
| Person-4 | 36-50 | Female | Major Heart issue, Obese |

| Person-5 | 50+ | Male | Lung Issue |

*3.2 Design Prototype and View of Real-Time Data*

After connecting all of the sensors to the Arduino and uploading the code to the board, the values of all the parameters appear on the screen and in the mobile application. The system works successfully to monitor sleep apnea. Fig.17 shows the hardware part of the prototype of the system.

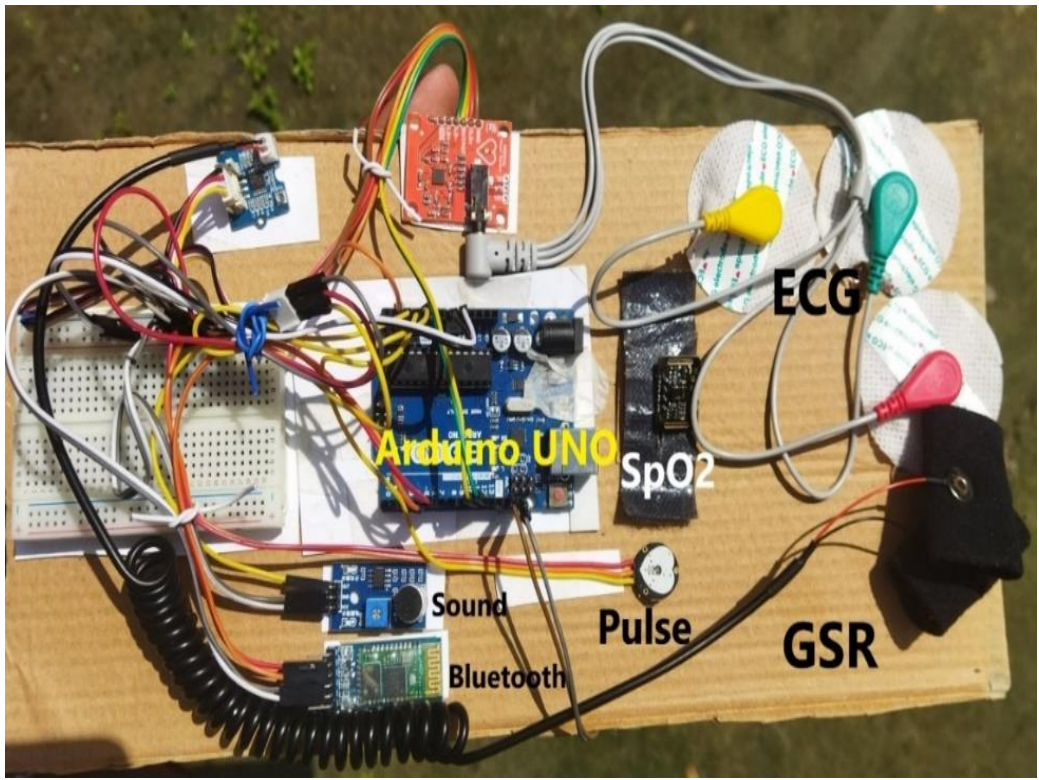

**Figure 17**: Prototype of the System.

The prototype system is integrated with the Arduino, sensors, and the Bluetooth module. Five different sensors are integrated in the system. Those sensors are marked in Fig. 17. The system includes a GSR sensor to measure the skin response, a pulse sensor to measure the heart rate, a Max30102 sensor to measure oxygen saturation (SpO2), an ECG sensor to monitor the electrocardiograms of the heart, and a sound sensor to measure the snoring intensity integrated into the system. The HC-05 Bluetooth module is in the system to perform serial communication between the hardware part and the mobile application.

Fig. 18 shows the prototype of the system during monitoring of a patient. The prototype of the system is an easily wearable device. It doesn't hamper the sleep of the patient. The application shown in Fig. 19 shows the real-time results of oxygen saturation (SpO2), heart rate, galvanic skin response, and sound intensity level. The mobile application can show the ECG graphs of the patients. But the serial plotter of the Arduino gives better ECG results.

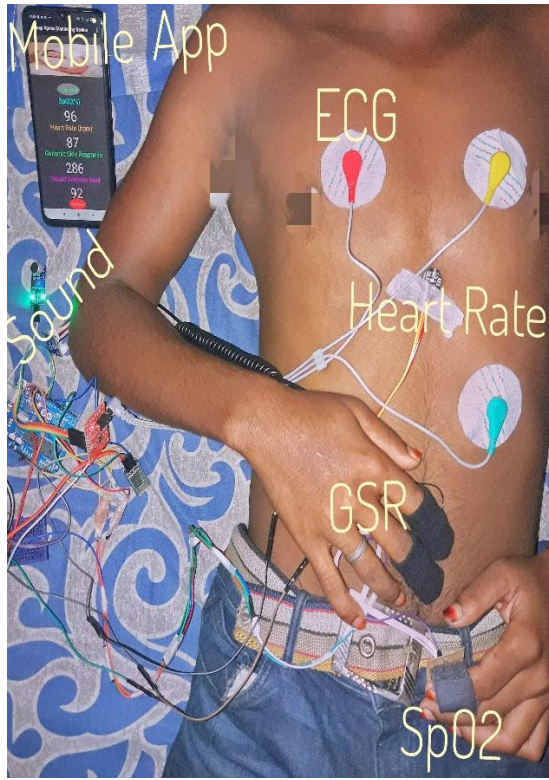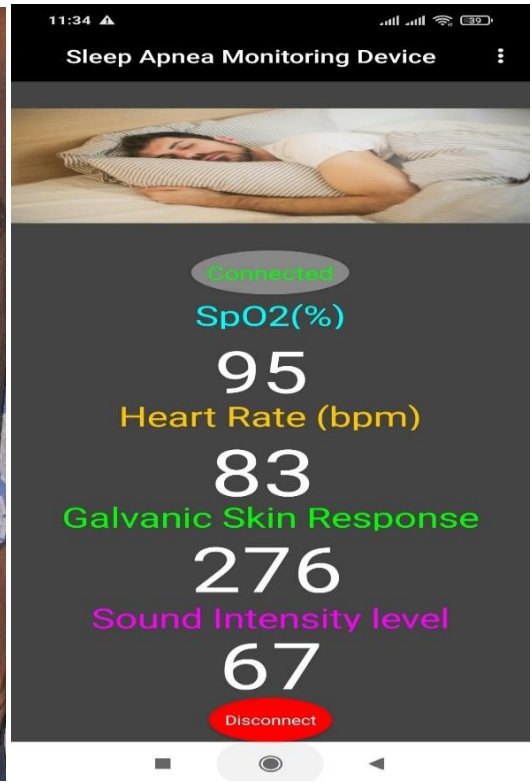

**Figure 18:** Sleep Apnea monitoring on a Patient    **Figure 19:** Mobile Application showing Real-time Result

*3.3 Data Sheet and Analytical Charts for Different Parameters*

In the experiment, the system monitors the person's sleeping period from some different perspectives and tries to see how the body's condition changes during the sleeping period. For that, it monitored the patient for six straight hours and analyzed how different parameters changed every hour.

*3.3.1 Heart Rate Analysis*

Tab.3 shows the results of the heart rate for the five people. It shows how the heart rates of different people change over hours. The values indicate how the heart rate of a person changes from the start of sleep to the end of sleep. In the experiment, we measured the value of heart rate every second throughout the period and checked the highest, lowest, and value that clicked most of the time. After checking the values, we took the decision to calculate the average heart rate in each hour. The average value reflects the heart rate that the person had at their maximum time in that specific hour. In the experiment for every person, the values are in some specific pattern, and in all cases, they follow the body condition of the person. For Person 1, who doesn't have any major health issues, has a very balanced heart rate throughout the sleeping period. The results don't provide any indication of sleep apnea. The same goes for Person 3, who is an athlete and has very good physical condition. His heart rate is in the range of 67-76 beats per minute during the period. That value is quite standard for a healthy person. But, for other people, the results do show some of the major hints about sleep apnea. Person 2 has an uneven change in heart rate. This person has a heart rate of around 90 bpm. In the third hour of sleep, his heart rate falls. One of the most serious consequences of breathing difficulties is a low heart rate, also known as Bradycardia. In that case, that low heart rate indicates obstructive sleep apnea in that patient. On the other hand, people 4 and 5 have a higher heart rate. According to the history, Person 4 has a major heart problem, which suggests that a higher heart rate could be the cause of both the heart problem and obesity.

**Table 3:** Result of BPM

| Hour | Person 1 | Person 2 | Person 3 | Person 4 | Person 5 |
| --- | --- | --- | --- | --- | --- |
| 1st Hour | 73.01 | 88.67 | 69.37 | 100.24 | 74.32 |
| 2nd Hour | 76.39 | 88.63 | 68.17 | 98.48 | 82.67 |
| 3rd Hour | 67.33 | 62.73 | 67.29 | 97.12 | 87.23 |
| 4th Hour | 72.82 | 83.21 | 75.87 | 79.54 | 89.56 |
| 5th Hour | 78.38 | 91.04 | 67.5 | 82.96 | 76.98 |
| 6th Hour | 78.87 | 85.34 | 73.68 | 90.02 | 73.57 |
| **Mean BPM** | **74.47** | **83.27** | **70.31** | **91.39** | **80.72** |

The average value of BPM of test subjects varied from person to person for different ages, physical conditions and previous health issues. There is quite a massive variation in the BPM of different persons.

For comparison of heart rate among the people, Fig. 20 shows how heart rate changes because of the physical condition of the people.

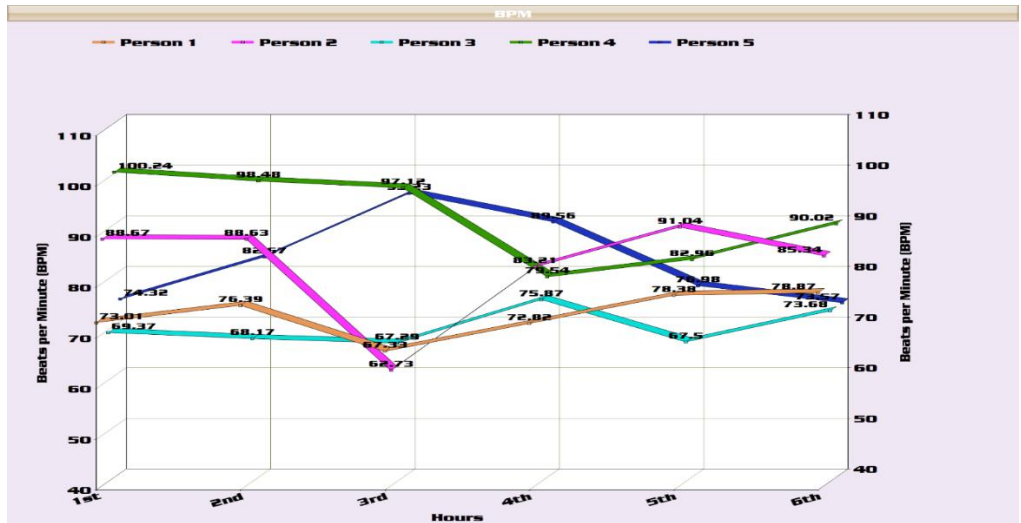

**Figure 20**: Line-chart for Heart Rate

The line chart of heart rate in Fig. 20 shows the average heart rate of a person changing over the course of an hour. In the line chart, it is visible that most of the people have an uneven change in heart rate. The differences between the lines of one person and another person indicate the difference in their age group and body condition.

*3.3.2 SpO2 Analysis*

Tab.4 shows the oxygen saturation for the five people. It shows the change of oxygen saturation over hours. In the experiment, the system calculates the SpO2 value every second with the sensor-based pulse oximeter. The SpO2 value always stayed within a particular range for all the people. For that reason, we took the average SpO2 value for every hour of the sleeping period. All of the people have quite good oxygen saturation levels throughout the monitoring period, except for Person 5, who has very low oxygen saturation during the third and fourth hours of sleep. That is one of the symptoms of breathing problems. A low rate of breathing declines the rate of oxygen supply to the body and it decreases the level of oxygen saturation. That low SpO2 value is an indication of obstructive sleep apnea.

**Table 4:** Result of SpO2

| Hour | Person 1 | Person 2 | Person 3 | Person 4 | Person 5 |
| --- | --- | --- | --- | --- | --- |
| 1st Hour | 95.56 | 95.90 | 97.12 | 95.71 | 95.87 |
| 2nd Hour | 95.97 | 95.79 | 98.11 | 96.45 | 94.15 |
| 3rd Hour | 95.68 | 95.96 | 97.70 | 96.73 | 93.78 |
| 4th Hour | 95.83 | 96.12 | 97.22 | 96.46 | 93.63 |
| 5th Hour | 96.04 | 96.00 | 98.37 | 97.49 | 94.12 |
| 6th Hour | 95.81 | 95.94 | 98.36 | 96.07 | 95.73 |
| **Mean SpO2** | **95.82** | **95.95** | **97.81** | **96.49** | **94.55** |

The average value of oxygen saturation for different people shown in Tab. 4 shows some variations. Although the variation is not significant, for Person-5 the value is a little lower and for Person-3 the value is higher than the other.

The line chart for SpO2 in Fig. 21 shows how the level of oxygen saturation changes over time more precisely.

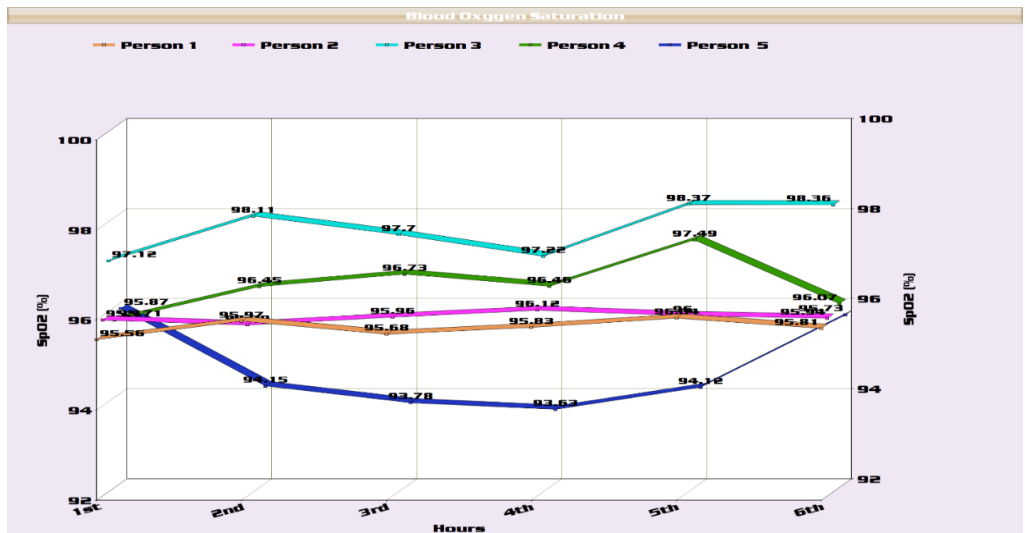

**Figure 21:** Line-chart for Oxygen Saturation (SpO2)

Line chart for the five persons in the Fig. 21 shows the change of oxygen saturation during different hour of sleeping period. It reflects how oxygen distribution happens in the body of those five people over the hours.

*3.3.3 Galvanic Skin Response (GSR) Analysis*

Galvanic Skin Response (GSR) values for the five people are shown in Tab.5. It shows every person's electro dermal activity throughout the sleeping period. The value indicates how a person's state of mind changes over time during sleep because of stress or hypertension. During the experiment, the system monitors the person's electro dermal activity via the sensor and shows the response of the skin every second. The response of the skin changes gradually for every person. For Person-1, the response of the skin is quite balanced throughout the sleeping time and it doesn't show any hint of sleep apnea, and also for Person-3, as his skin response is highest among the others. But for person 2, person 4, and person 5, the result shows an uneven change in skin response. For Person 2, the value of GSR goes down in the 3rd hour and also in the last two hours. It shows the chance of stress and hypertension problems during sleep. Person 4 has the lowest GSR value among all of the people the system monitored. In the first two hours, the values of skin response don't show any change, which can lead to a decision about sleep apnea. But from the 3rd hour to the end of the sleep, the GSR value reflects the change of very low dermal activity, which can be the reason for sleep apnea. Person 5 also has quite a low skin response, but the value throughout the time doesn't change significantly, and in that case, there is no evidence for sleep apnea for person 5.

**Table 5**: Result of GSR

| Hour | Person-1 | Person-2 | Person-3 | Person-4 | Person-5 |
|---|---|---|---|---|---|
| 1st Hour | 231.03 | 181.81 | 348.2 | 188.88 | 190.54 |
| 2nd Hour | 235.47 | 254.87 | 269.35 | 150.03 | 216.47 |
| 3rd Hour | 264.67 | 134.34 | 265.41 | 74.22 | 142.79 |
| 4th Hour | 221.74 | 154.95 | 217.13 | 123.25 | 135.38 |
| 5th Hour | 193.13 | 123.82 | 253.77 | 119.51 | 147.90 |
| 6th Hour | 207.18 | 125.21 | 298.82 | 118.08 | 164.27 |
| **Mean GSR** | **225.54** | **162.5** | **275.45** | **128.95** | **166.23** |

The mean of galvanic skin response for five people in Tab. 5 shows significant variations in the skin response over the hour.

The line chart shows the change in Galvanic Skin Response for the five people over the hour in Fig. 22 more precisely.

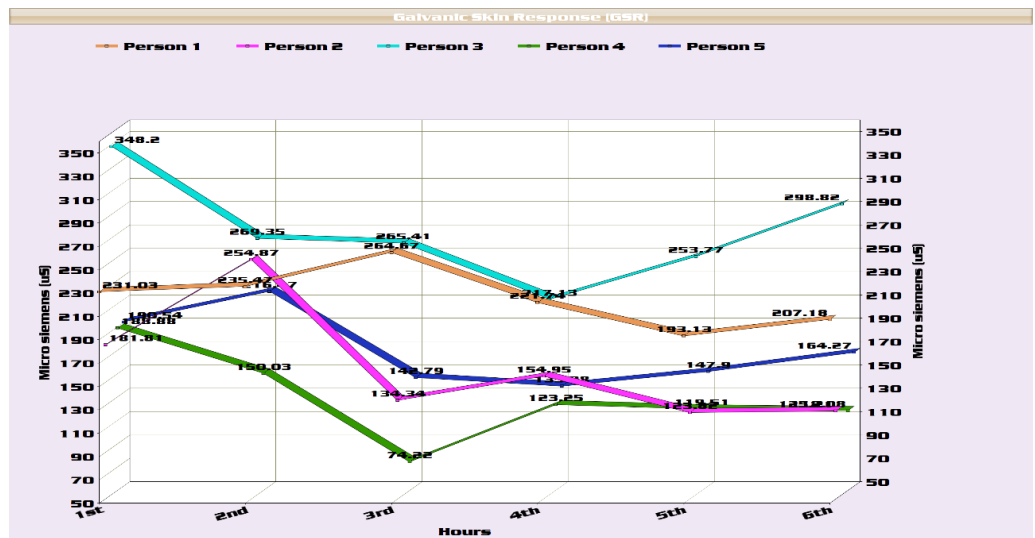

**Figure 22**: Line-chart for Galvanic Skin Response

*3.3.4 Sound Intensity Analysis*

Snoring, which is a symptom of obstructive sleep apnea, was detected in three of the people the system monitored. The snoring sound followed a pattern for each person. In the experiment to detect snoring, the sensor collects data every second. The result shows that, for a specific person, the sound follows a similar rhythm throughout the sleeping time. The graph in Figs. 23 (a-e) depicts the snoring rhythms of the people studied.

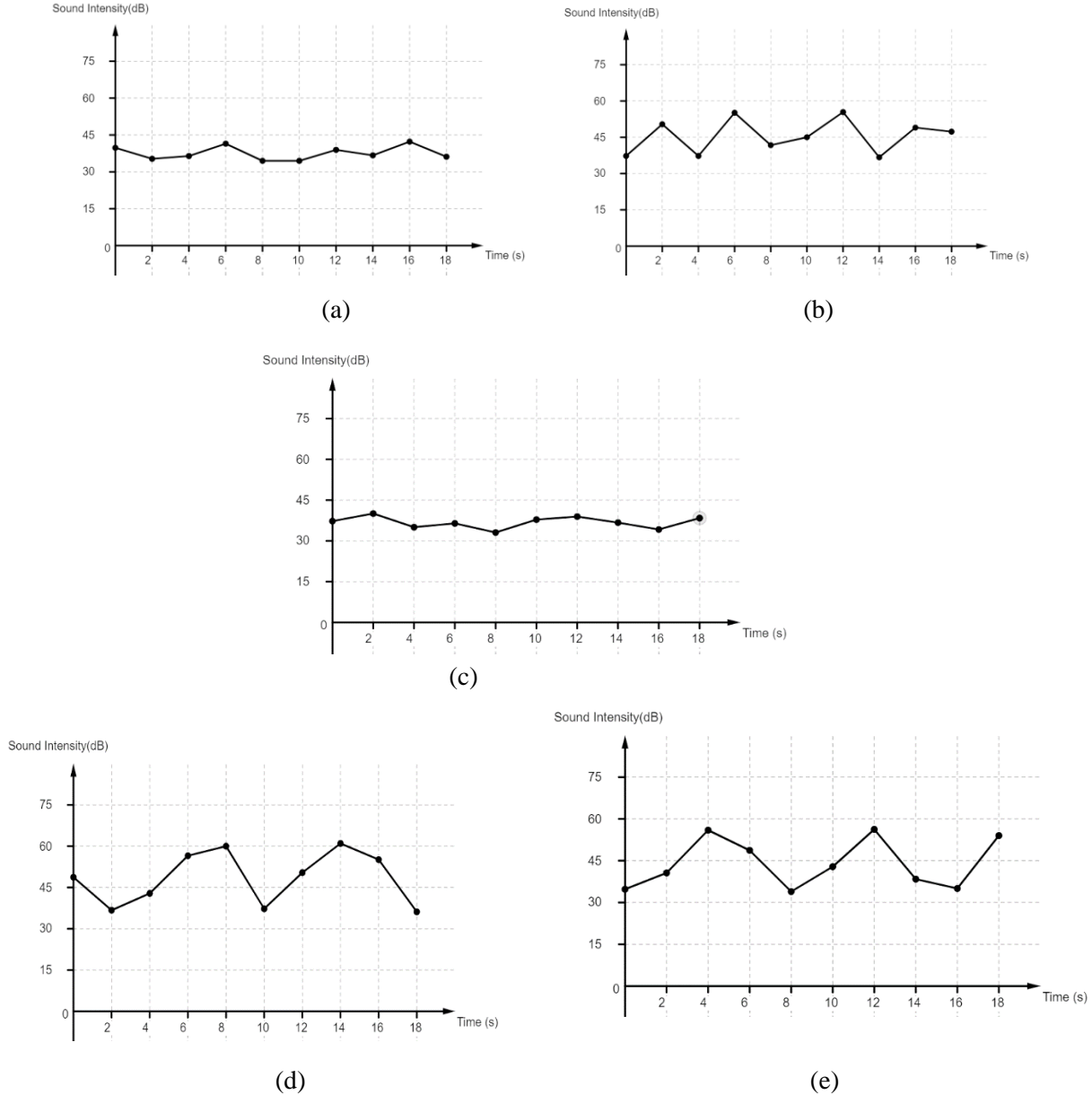

**Figure 23:** Snoring Cycle of 5 Persons; (a), (b), (c), (d) and (e) shows the Sound Intensity of Person 1, Person 2, Person 3, Person 4 and Person 5 respectively.

Person 3 has a very balanced rhythm in the breathing cycles. There is no evidence of snoring visible in Fig. 23 (c). The sound level stayed in the range of 35 dB - 40 dB, which is usually for room sound. But for person 4 and person 5, there is a big problem with snoring, as it is visible in Fig. 23 (d) and Fig. 23 (e). For person-4, the figure shows that the snoring sound reached almost 60 dB and every snoring cycle took 3-5 seconds. That is big evidence of problems with breathing and also a sign of obstructive sleep apnea. The same goes for person 5, as snoring intensity touches the 60 dB mark. The breathing rate is lower than the normal rate as the breathing cycle for person-5 took 4-6 seconds. Both people faced problems with breathing during sleep, which is a sign of sleep apnea.

*3.3.5 ECG Analysis*

Fig. 24 depicts the heart rhythm or ECG graph for test person 4. Person 1, person 2, person 3 and person 5 have a very balanced ECG result. But the ECG graphs of person 4 show some uneven changes in the rhythm of the heart.

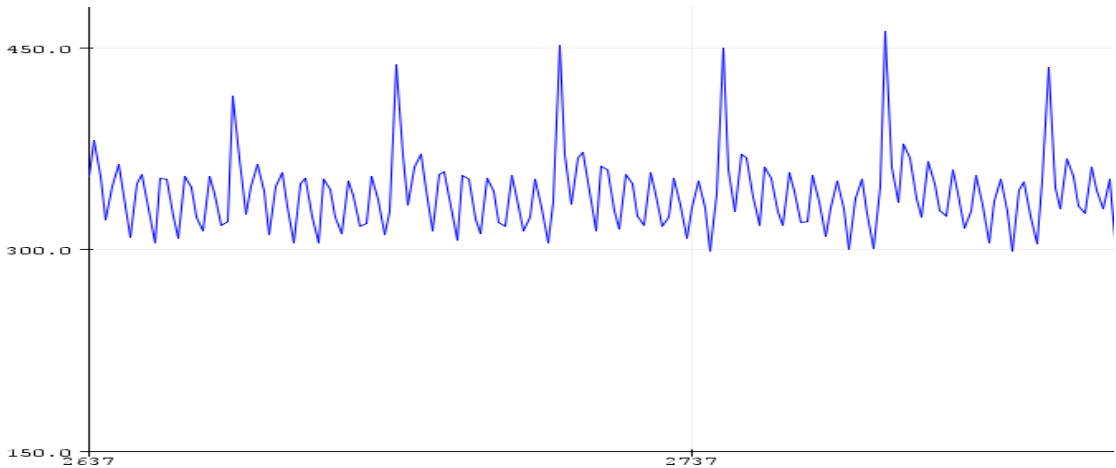

**Figure 24**: ECG Graph of Test Person 4

In Fig. 24, there are six waves that are the top picks of the graph. Those spikes are also known as R-wave. It shows how time elapsed between endocardium to epicardium. It is also helpful for some of the major heart issues. Loss of R-wave indicates myocardial infarction, which is visible in the first R-pick in Fig. 24. The gain of R-wave indicates cardiomyopathy, which is the 5th R-pick of the graph.

*3.4 Comparative Analysis*

The outcome of the system has been compared to some other research that has been monitoring sleep apnea with similar kinds of parameters shown in Tab. 6. It also shows the parameters that weren't used in other research.

**Table 6**: Comparison Table with Previous Studies

| Reference | Main parameters | Result (final or mean) |
|---|---|---|
| In the paper [25] | Oxygen Saturation (SpO2) | $(91.3 \pm 2.0)$ % <br> The oxygen saturation for sleep apnea patient is in this range. |
| In the paper [6] | Heart Rate | $(70.9 \pm 2.2)$ bpm <br> The study detected sleep apnea in this range of heart rate. |
| In the paper [26] | Polysomonography | 91.8% to 96.5% accuracy |

|  |  | on adaptive Apnea index |
| --- | --- | --- |
| In the paper [27] | Snoring Sound Intensity | 45 dB to 58 dB and mean value is approx. 52 dB. |
| In the paper [11] | 3D Camera | Novel approach proposed to measure respiratory movement using 3D camera to detect Apnea. |
| Our system | SpO2, Heart Rate, ECG, Skin Response, Snoring Sound Intensity | SpO2: 94.55 to 97.81<br>BPM: 74.47 to 91.39<br>GSR: 128.95 to 275.45<br>Sound: 35 dB to 60 dB<br>ECG: generalized |

Most of the previous research to monitor sleep apnea is mainly based on one parameter. But this study combined five different parameters of the sleeping index. In this study, the decision was made after analyzing all the five parameters combined as one factor is dependent on another factor of sleep. Another important aspect of this study is no other research focused on the Galvanic Skin Response to detect sleep apnea. However, this study establishes a link between skin response and sleep apnea monitoring. That opens an important area for analyzing sleep apnea more precisely.

**4 Conclusion**

This research shows how IoT devices can monitor sleep apnea. To implement the system, we used a basic microcontroller and some of the major health-related sensors. The mobile application has been created with a very simple app developing web application. After monitoring five people, the system gives quite satisfactory results for making decisions about sleep apnea. From the given results, it is clear that two people do not have any symptoms of any kind of sleep apnea. One person who is in the age range of 36-50 has major issues with sleep conditions. The system successfully detects sleep apnea for that person. The system also detects obstructive sleep apnea in a person. After analyzing the results, it is clear that the person whose age is 50+ is a patient with OSA. That kind of monitoring will help people to detect sleep apnea at the early stage. Thus, this research can help people to learn about sleep apnea, the way to detect it, and it will also help people to eliminate all their sleeping problems.

Nowadays, sleeping disorder rates are higher than before. More people are suffering from hypertension, stress related problems and heart related problems which cause sleep apnea. Our system will be very much more sustainable to monitor sleep apnea and detect specifically the duration of sleep apnea. That system can sustain itself at a higher level. It not only detects sleep apnea but can also inform the patient to discuss with the doctor the data taken from the system which can make the lives of patients easier. And the project can be sustained in the future if we add higher technology like high-level X-rays and other monitoring systems. Some other sensors could be added to the system to make it more applicable. An LM35 temperature sensor could be added to the system. A DHT11 humidity sensor could also be added to this system to make this system more applicable, though those factors have a minor connection with sleep apnea.

In most research, ECG data and other data are used to measure or monitor sleep apnea. Most were under medical observation in a clinical environment. But our device will help patients to get even better output while staying at home and with ease of use. Real-time data will give instant updates about apnea indexes so that patients can take precautionary steps. And for senior citizens and the physically/mentally disabled, our device is an added benefit because it sends notification data to the patient's concerned person so that the person can take care of any difficulties that may occur at any time. We will further try to build our own dataset with this real-time data and will analyze it through multiple Machine Learning models to analyze the whole system and how our system can work with better efficiency. To have a healthy and sound life without a sleeping disorder is a blessing, and technology can undoubtedly assure that.

**Data Availability Statement:** No data was utilized to support this research findings.

**Funding Statement:** The authors are thankful for the support from Taif University Researchers Supporting Project (TURSP-2020/26), Taif University, Taif, Saudi Arabia.

**Conflicts of Interest:** The authors declare that they have no conflicts of interest to report regarding the present study.